\documentclass{nature}
\usepackage{amsmath,amssymb,epsfig,epsfig,ifthen,graphicx}
\usepackage{multibib}
\usepackage{subcaption}

\newcommand\aj{Astron. J.}
\newcommand\araa{Ann. Rev. Astron. Astrophys.}
\newcommand\apj{Astrophys. J.}
\newcommand\apjl{Astrophys. J. Lett.}      
\newcommand\apjs{Astrophys. J. Suppl.}
\newcommand\aap{Astron. \& Astrophys.}
\newcommand\mnras{Mon. Not. R. Astron. Soc.}
\newcommand\nat{Nature}
\newcommand\pasa{Publ. Astron. Soc. Aus.}

\usepackage{aas_macros}
\usepackage{xspace}    
\usepackage{microtype} 
\usepackage{float}

\begin{document}

\newcommand{\um}{\ensuremath{\mu\mathrm{m}}\xspace}
\newcommand{\uJy}{\ensuremath{\mu\mathrm{Jy}}\xspace}
\newcommand{\degrees}{$^{\circ}$}
\newcommand{\degsq}{\ensuremath{\mathrm{deg}^2}}
\newcommand{\ssfr}{\ensuremath{\mathrm{sSFR}}\xspace}
\newcommand{\ssfrunit}{\ensuremath{\mathrm{yr}^{-1}}}
\newcommand{\oii}{[\textrm{O}~\textsc{ii}]}
\newcommand{\mgii}{\textrm{Mg}~\textsc{ii}}
\newcommand{\mgi}{\textrm{Mg}~\textsc{i}}
\newcommand{\neiii}{[\textrm{Ne}~\textsc{iii}]}

\def\mpch {$h^{-1}$ Mpc} 
\def\kpch {$h^{-1}$ kpc} 
\def\kms {km s$^{-1}$} 
\def\lcdm {$\Lambda$CDM } 
\newcommand{\hMpc}{\ensuremath{\,h^{-1}\, \textrm{Mpc}}}
\def\etal {et al.}
\def\kt{\tilde{k}}
\def\mpc{{\rm Mpc}}

\def\msunyr{\hbox{$M_{\odot}~{\rm yr}^{-1}$}}
\def\deg{\hbox{$^{\circ}$}}

\newcommand\arcsec{\mbox{$^{\prime\prime}$}}
\newcommand{\mass}{\ensuremath{\mathrm{M_*}}}
\newcommand{\mstar}{\ensuremath{\mathcal{M}_*}}
\newcommand{\msun}{\ensuremath{M_{\odot}}}
\newcommand{\logmass}{\ensuremath{\log\,(\mass/\msun)}}
\newcommand{\sfr}{\ensuremath{\psi}}
\newcommand{\sfrm}{{\rm SFR}/\ensuremath{\mass}}
\newcommand{\sfrunit}{\ensuremath{\mathcal{M}_{\sun}~\textrm{yr}^{-1}}}
\newcommand{\arcsecsq}{\ensuremath{\mathrm{arcsec}^2}\xspace}

\title{Ionized Gas Extended Over 40 kpc in an Odd Radio Circle Host Galaxy}

\author{
  Alison L. Coil$^{*1}$,
  Serena Perrotta$^1$,
  David S.~N. Rupke$^2$,
  Cassandra Lochhaas$^3$,
  Christy A. Tremonti$^4$,
  Aleks Diamond-Stanic$^5$,
  Drummond Fielding$^6$
  James Geach$^7$,
  Ryan C. Hickox$^8$,
  John Moustakas$^9$,
  Gregory H. Rudnick$^{10}$,
  Paul Sell$^{11}$,
  Kelly E. Whalen$^8$
}

\maketitle

\begin{affiliations}
\item Center for Astrophysics and Space Sciences, University of California, 9500 Gilman Dr., La
  Jolla, CA 92093, acoil@ucsd.edu
\item Department of Physics, Rhodes College, Memphis, TN 38112
\item Space Telescope Science Institute, 3700 San Martin Dr., Baltimore, MD 21218
\item Department of Astronomy, University of Wisconsin-Madison, Madison, WI 53706
\item Department of Physics and Astronomy, Bates College, Lewison, ME 04240
\item Center for Computational Astrophysics, Flatiron Institute, 162 5th Ave, New York, NY 10010
\item Centre for Astrophysics Research, University of Hertfordshire, Hatfield, Hertfordshire AL10 9AB
\item Department of Physics and Astronomy, Dartmouth College, Hanover, NH 03755
\item Department of Physics and Astronomy, Siena College, Loudonville, NY 12211
\item Department of Physics and Astronomy, University of Kansas, Lawrence, KS 66045
\item Department of Astronomy, University of Florida, Gainesville, FL, 32611
\end{affiliations}

\begin{abstract}
  A new class of extragalactic astronomical sources discovered in
  2021, named Odd Radio Circles (ORCs)\cite{Norris21a}, are
  large rings of faint, diffuse radio continuum emission spanning
  $\sim$1 arcminute on the sky. Galaxies at the centers of several ORCs
  have photometric redshifts of $z\sim0.3-0.6$, implying physical scales
  of several 100 kiloparsecs in diameter for the radio emission, the origin
  of which is unknown.
  Here we report spectroscopic data on an ORC including
  strong \oii \ emission tracing 
  ionized gas in the central galaxy of ORC4 at $z=0.4512$.
  The physical extent of the \oii \ emission is $\sim$40 kpc
  in diameter, larger than expected for a typical early-type
  galaxy\cite{Pandya17} but an order of magnitude smaller than the
  large-scale radio continuum emission.  We detect a $\sim$200
  \kms \ velocity gradient across the \oii \ nebula, as well as a high
  velocity dispersion of $\sim$180 \kms.  The \oii \
  equivalent width (EW, $\sim$50 \AA) is extremely high for a quiescent
  galaxy.  The morphology, kinematics,
  and strength
  of the \oii \ emission are consistent with the infall of shock ionized gas near the galaxy, following a
  larger-scale, outward moving shock driven by a galactic wind. Both the
  extended optical and radio emission, while observed on very different scales,
  may therefore result from the same dramatic event.

\end{abstract}

Three ORCs were first discovered\cite{Norris21a} in
the Evolutionary Map of the Universe Pilot
Survey\cite{Norris21b} at a frequency of $\sim$1 GHz using the
Australian Square Kilometer Array Pathfinder (ASKAP)
telescope\cite{Johnston07,McConnell16}.  The ASKAP array has high 
angular resolution ($\sim$13\arcsec), is sensitive to low surface
brightness emission, and the pilot survey covers 270 deg$^2$ of the
sky, such that it is able to detect rare, faint objects not
previously detected.  A fourth ORC (ORC4) was discovered in
archival data taken with the Giant MeterWave Radio Telescope
(GMRT)\cite{Anantha95} at 325 MHz, while additional ORCs were discovered in
later ASKAP\cite{Koribalski21} and MeerKAT data\cite{Koribalski23}.
ORCs exhibit large, limb-brightened rings of radio continuum emission
with lower surface brightness emission in the interior.
Follow-up MeerKAT data on ORC1 reveals
that the radio emission is most likely
aged synchrotron, and polarization data show magnetic field lines that
are consistent with an expanding shell\cite{Norris22}.
The origin of the radio emission is unknown, with scenarios proposed
including Galactic supernovae remnants, double-lobed radio galaxies,
ring galaxies, and gravitationally-lensed Einstein rings\cite{Norris21a}.
However, the most likely scenarios involve a shock from an outflowing
galactic wind or a blast wave driven by merging supermassive black
holes\cite{Koribalski21}.

To search for an  optical counterpart to ORCs, we observed ORC4
with the Keck Cosmic Web Imager (KCWI\cite{Morrissey18})
and detected strong \oii \ and weak \mgii \  and \neiii \ line emission
in a spatially-integrated spectrum (Extended Data Fig.~1).
The \oii \ luminosity is $7.1 \times 10^{41} \ erg \ s^{-1}$, near the break
of the \oii \ luminosity function at $z\sim0.5$\cite{Zhu09}.  The restframe \oii
\ EW is 50 \AA \ in the spatially-integrated spectrum, which is an order of
magnitude higher than what is typically found in red, early-type galaxies\cite{Yan06}.
Long-slit optical spectroscopy of ORC4 that
includes wavelengths redward of the KCWI data reveals low-ionization
nuclear emission-line region (LINER)-like line ratios (D.~S.~N. Rupke
et al., manuscript in preparation).

A spectral energy distribution (SED) fit to the optical and infrared
emission in the central galaxy of ORC4 (Extended Data Fig. 2) 
reveals that the galaxy is massive (log $M_*/M_{\odot} = 11.27 \pm0.06$)
and has an old stellar population, with a  stellar age of $t_{age} = 6.0 \ (+/-1.6)$ Gyr.
While the SED fit has only a 3\% contribution from an
AGN, the radio continuum emission in this galaxy\cite{Norris21a}
is likely due to an AGN, as there is no indication of on-going star
formation.  The observed
radio flux of 1.43 mJy at 325 MHz corresponds to a luminosity of
1.24$\times10^{24} \ W \ Hz^{-1}$ at a restframe frequency of 200 MHz,
given the redshift and radio spectral index\cite{Norris21a}; this is
1.5 orders of magnitude fainter than the break in the radio AGN
luminosity function\cite{Franzen21}.

The \oii \ emission is bright enough to create spatially-resolved maps
 to probe the morphology, extent, and resolved
kinematics of the ionized gas.  Fig.~1 shows a map of the \oii
\ surface brightness, the peak of which is $2.3 \times 10^{-16} \ erg
\ s^{-1} \ cm^{-2} \ arcsec^{-2}$.
The azimuthally-averaged surface brightness profile (Fig.~2) reveals a
strong central concentration of the \oii \ emission, in contrast to
the larger-scale limb-brightened radio emission.  The radial extent is $\sim$20 kpc
($\sim$40 kpc in diameter), and the half light (or ``effective'')
radius is 3.4 kpc, deconvolved with the seeing.
The \oii \ radial extent is an order of magnitude smaller than the
radio continuum emission in ORC4, which extends $\sim$200 kpc in radius.

The \oii \ restframe EW spatial map (Fig.~1, panel d) does not show
strong asymmetry.  The restframe EW at the
center of the nebula is $>$70 \AA (Fig.~2), and at a radius of
15 kpc (shown here at $R/R_e = 3$, relative to the galaxy half-light radius)
it is $\sim$25 \AA.  For comparison, \oii \ EW profiles are shown for
massive, nearby early-type galaxies, where integral field data reveals
that 38\% of galaxies with $>10^{11.5}$~M$_{\odot}$ have \oii
\ emission\cite{Pandya17}.  The median \oii \ EW of these comparison
galaxies is 4.8 \AA, and the maximum EW is 28 Ang.
The \oii \ EW in ORC4 is therefore an order of magnitude higher than is typically seen 
in massive, early-type galaxies.
The maximal radial extent of the \oii \ in these comparison galaxies
varies from 0.6-18.2 kpc, with a median maximal extent of 2.3 kpc.
The \oii \ emission in ORC4, detected to a radius of $\sim$20
kpc, is therefore unusually spatially extended.  It is extended not only
in the sense that it is not from the nucleus of the galaxy, but the
maximal extent is an order of magnitude higher than in comparable
galaxies.

The kinematics of the \oii \ emission (Fig.~3) reveal an asymmetric
velocity gradient across the nebula, from $+$170 \kms at the northern edge to
$-50$ \kms at the southern edge. 
A north-south position-velocity diagram (Fig.~3, panel c) highlights the asymmetry in the
velocity gradient.  The velocity dispersion varies
from a minimum of $\sim$60 \kms \ at the edges of the nebula to
$\sim$180 \kms \ in the center and up to $\sim250$ \kms \ at the
northern and western edges.

There are multiple possibilities for the origin of the extended \oii
\ emission observed in ORC4. Given the properties of the ionized gas and
host galaxy, it is unlikely that it is associated with the interstellar
medium (ISM) of the galaxy or an AGN or AGN-driven outflow or due to a cooling flow
(see Methods section ``Alternative scenarios for ionized gas origin'' for details.)
Another possibility is that it is associated with
the same energetic event that created the large-scale radio emission
which identified the source as an ORC. The LINER-like line ratios of ORC4
point to the ionized gas arising from shocks\cite{Yan12}.  One of the possible 
scenarios for the creation of ORCs is a shock from a starburst wind, which
would create multiple shocks on different scales including both a forward
moving shock on large scales and a reverse shock, sometimes called a wind shock,
on smaller scales, with
a contact discontinuity in between\cite{Faucher-Giguere12,King15}.
In a model of galactic winds created by starburst
activity\cite{Thompson16}, the forward shock can exist on scales of
$>$100 kpc, while the reverse shock exists on scales of $\sim$10-50
kpc.  Once the starburst episode that initially drove the outflow has
shut off, the gas heated by the reverse shock can then travel
backwards towards the central galaxy\cite{Lochhaas18}, moving to 
smaller scales. The shocked wind expands to fill the under-pressurized
inner region that has been cleared by the outward-moving forward shock
and may produce a turbulent, energetic medium full of additional
shocks by interacting with gas in and around the galaxy.  Wind
re-heated in this way by the reverse shock
may then radiatively cool giving rise to ionized gas
emission on the scale of the galaxy.  Meanwhile the forward shock,
which has continued to propogate outwards, creates synchrotron
emission on large scales.
This scenario differs than that proposed by
Norris et al. (2022), where the large-scale radio emission is
associated with the reverse shock; we propose that is
associated with the forward shock.
In theoretical models the large-scale
forward shock can propogate to scales of hundreds of kiloparsecs on
timescales of several Gyr\cite{Lochhaas18}.
Recent cosmological simulations focusing on
galaxy mergers as the cause of 
ORCs\cite{Dolag23} show that additional shocks are seen on smaller scales as well.

This scenario is consistent with observations of ORC4 in
multiple ways.  The best fit SED model indicates that the galaxy
had a burst of star formation $\sim$1 Gyr ago and is no longer forming stars.
The scale, morphology, and surface brightness distribution of the \oii \ emission
are consistent with expectations for the shocked wind gas falling back towards
the galaxy, which could also lead to an asymmetric velocity gradient and high
velocity dispersion due to turbulence.
The strength of the \oii \ emission is also consistent with being due
to shocks, which are known to lead to an enhancement of
\oii\cite{Allen08}, and/or may reflect mixing between hot and cool gas
phases, which can lead to additional gas cooling\cite{Gronke20,Fielding22}.

To further test a starburst-driven wind model for ORCs, we ran a
suite of simulations that launch a wind from an isolated
galaxy and followed the evolution of the forward and reverse shocks
(Fig. 4). Our goal was to reproduce the forward shock
radius ($\sim$200 kpc) and Mach number ($\sim$1-2) required to produce the observed
synchrotron emission, the scale of the reverse shock and subsequent infalling
shocked wind ($\sim$20 kpc), the elevated velocity dispersion of the shocked wind
gas ($\sim$150 km s$^{-1}$), and the age of the wind bubble ($\sim$1 Gyr).  We found that
these parameters could be reproduced with a wind that is launched with a
velocity of 450 km s$^{-1}$ and a mass outflow rate of 200 \msun
yr$^{-1}$, which blows for 200 Myr.
(See Methods section ``Outflowing galactic wind simulation'' for details.)
The high mass outflow rate needed is potentially consistent with the SED fit results for
the central galaxy in ORC4, which require a high fraction of the total stellar mass
(61\%) produced in a starburst which presumably had a high SFR, which could have created
an outflow with a high mass outflow rate. 
The features of the wind model presented here are generic features of
an energy and mass injection, without regard to the source of the
injection. In order for the wind bubble to survive at a Mach number of $\sim$1-2
to a scale of $\sim$200 kpc, the density of the CGM must be lower than is typically assumed
for a galaxy of stellar mass $10^{11}$ \msun, though the density is still feasible (simulations find the CGM density of a Milky Way-mass halo spans six orders of magnitude\cite{Simons20}).
As ORCs appear to be a rare
phenomena, they may arise only in situations where the galaxy progenitor
and CGM parameters are favorable. This simulation shows that a single event such as a strong
starburst can launch a wind that simultaneously produces both the
large-scale radio emission and the smaller-scale \oii \ emission
scale, when observed $\sim$1 Gyr after the event.

We have been studying outflowing galactic winds in
massive ($\sim10^{11}$~M$_{\odot}$) starburst galaxies at
$0.4<z<0.8$\cite{Tremonti07}; most are late-stage major mergers\cite{Sell14}.
These 
starburst galaxies are driving extremely fast multi-phase gas outflows
with velocities of $\sim$2000~km~s$^{-1}$ and mass outflow rates of
$\gtrsim$200 \msun yr$^{-1}$\cite{Rupke23, Perrotta23}.
Their space densities ($1 \times 10^{-6}
Mpc^{-3}$) are similar to that of post-starburst galaxies and
ultra-luminous infrared galaxies (ULIRGs)\cite{Whalen22}.
KCWI data reveals \oii \ nebulae in these galaxies of
sizes 30-100 kpc\cite{Rupke19} (S. Perrotta et al., in preparation).

Though their space density is potentially lower (estimated at $\sim2
\times 10^{-8} Mpc^{-3}$)\cite{Norris22},
ORCs may be a later stage of
a similar phenomenon.  In order for ORCs to be the result of a
starburst wind shock the starburst must have been
extreme\cite{Norris22}, with a high mass outflow rate, and the density of
the CGM must be low. 
The massive starburst galaxies we have been studying at $z\sim0.5$ have such
extreme outflow properties\cite{Diamond-Stanic21}.
In the model of an ancient, starburst-driven wind, the
ancestors of ORCs must resemble these compact starbursts in some
way. These ORC wind relics may then be a direct tracer of
powerful outflows from earlier energetic starburst events.

\clearpage

\begin{figure}
  \begin{center}
  \includegraphics[width=0.95\textwidth]{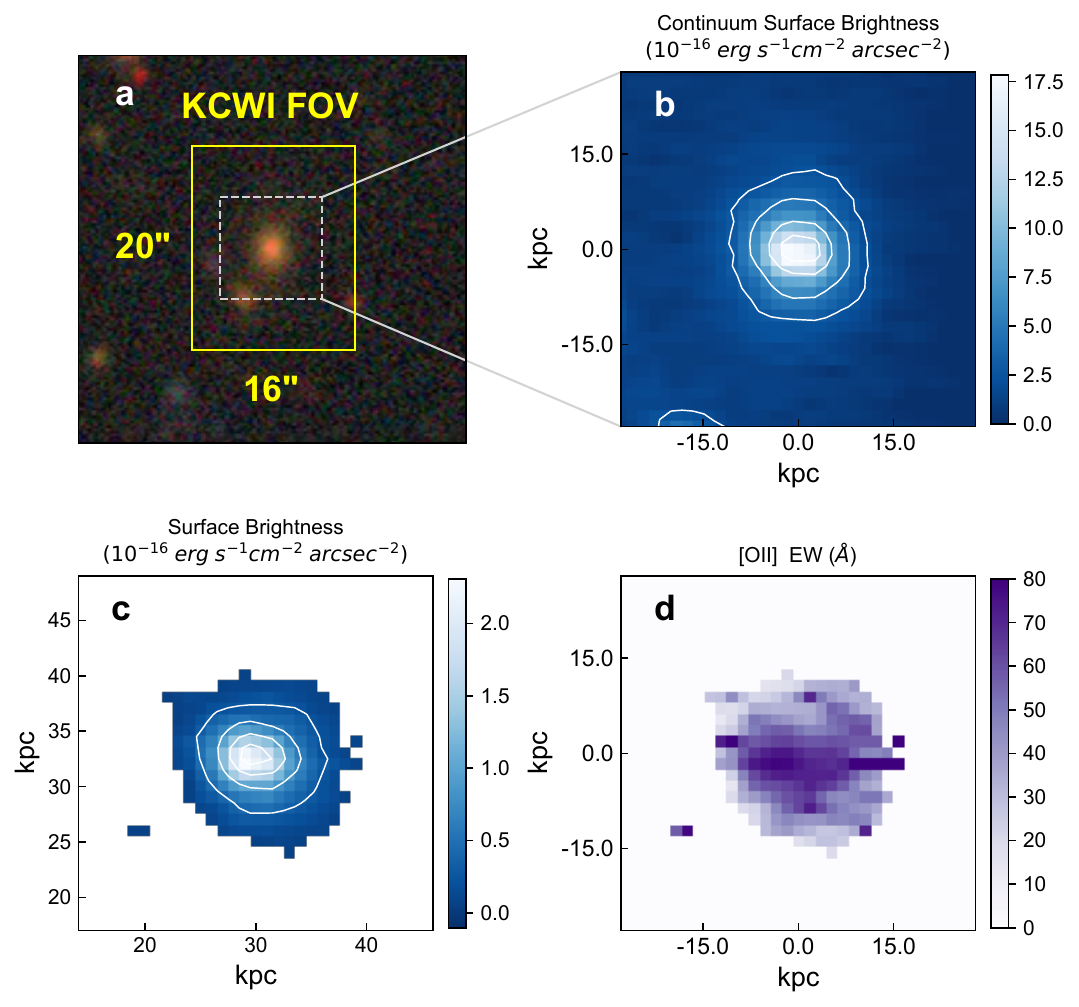}
  \end{center}
  \caption{\small {\bf Optical stellar continuum and \oii \ emission in ORC4.}
    (a) DESI Legacy Survey DR9\cite{Dey19} optical $grz$ image of ORC4,
    spanning 40$\arcsec$ on each side.
    The KCWI field of view (FOV) is shown with a yellow box, and the dotted
    white box traces the
    central 10\arcsec \ $\times$ 10\arcsec, shown in the remaining panels.
    (b) Stellar continuum surface brightness image from KCWI,
    integrated over the observed
    wavelength range 4200-5300 \AA, with axes labeled in kpc from the center
    of the galaxy.
    White contours show levels of
    2.5, 5, 10, and 15 $\times \ 10^{-16} \ erg \ s^{-1} \ cm^{-2} \ arcsec^{-2}$.
    (c)  \oii \ surface brightness image, showing spaxels
    with S/N$\ge$3.  White contours show surface brightness levels of
    0.3, 0.8, 1.4, and 1.9 $\times \ 10^{-16} \ erg \ s^{-1} \ cm^{-2} \ arcsec^{-2}$.
    (d) \oii \ restframe equivalent width (EW) map.
  }
  \label{OII_spatial}
\end{figure}

\clearpage

\begin{figure}
  \begin{center}
  \includegraphics[width=0.95\textwidth]{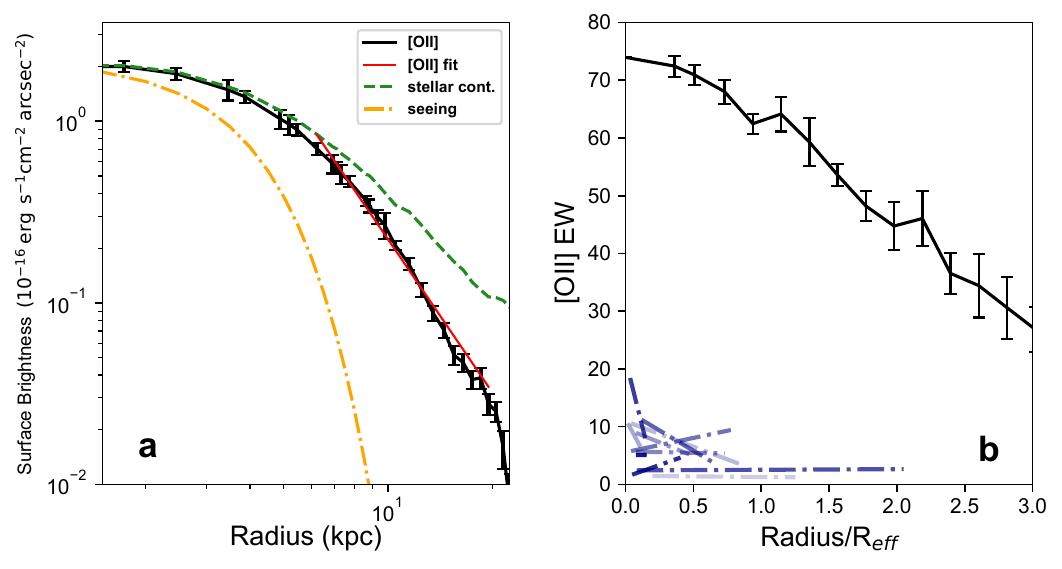}
  \end{center}
  \caption{\small {\bf Radial surface brightness and equivalength width profile of the \oii \ emission.}
    (a) The black line shows the \oii \ radial surface
    brightness profile, with 1$\sigma$ errors;
    the nebula has a radial extent of $\sim$20 kpc.  The green line shows
    the radial profile of the stellar continuum, normalized to the peak of
    the \oii \ profile, and the orange line shows the extent of the
    ground-based seeing, also normalized to the \oii.
    The red line is the best fit slope to the \oii \
    radial profile (beyond 5 kpc, outside the seeing disk) of $-2.8$.
    (b) The black line shows the restframe
    equivalent width (EW) profile of the \oii \
    emission in ORC4 as a function of radius from the center of the galaxy,
    relative to the stellar half light radius ($R_e=4.8$ kpc), with 1$\sigma$ errors.
    For comparison, purple dot-dashed lines show the \oii \ EW radial profile fits for
    nearby massive, early-type galaxies\cite{Pandya17}. 
  }
  \label{radial_profiles}
  \end{figure}

\clearpage

\begin{figure}
  \begin{center}
  \includegraphics[width=0.95\textwidth]{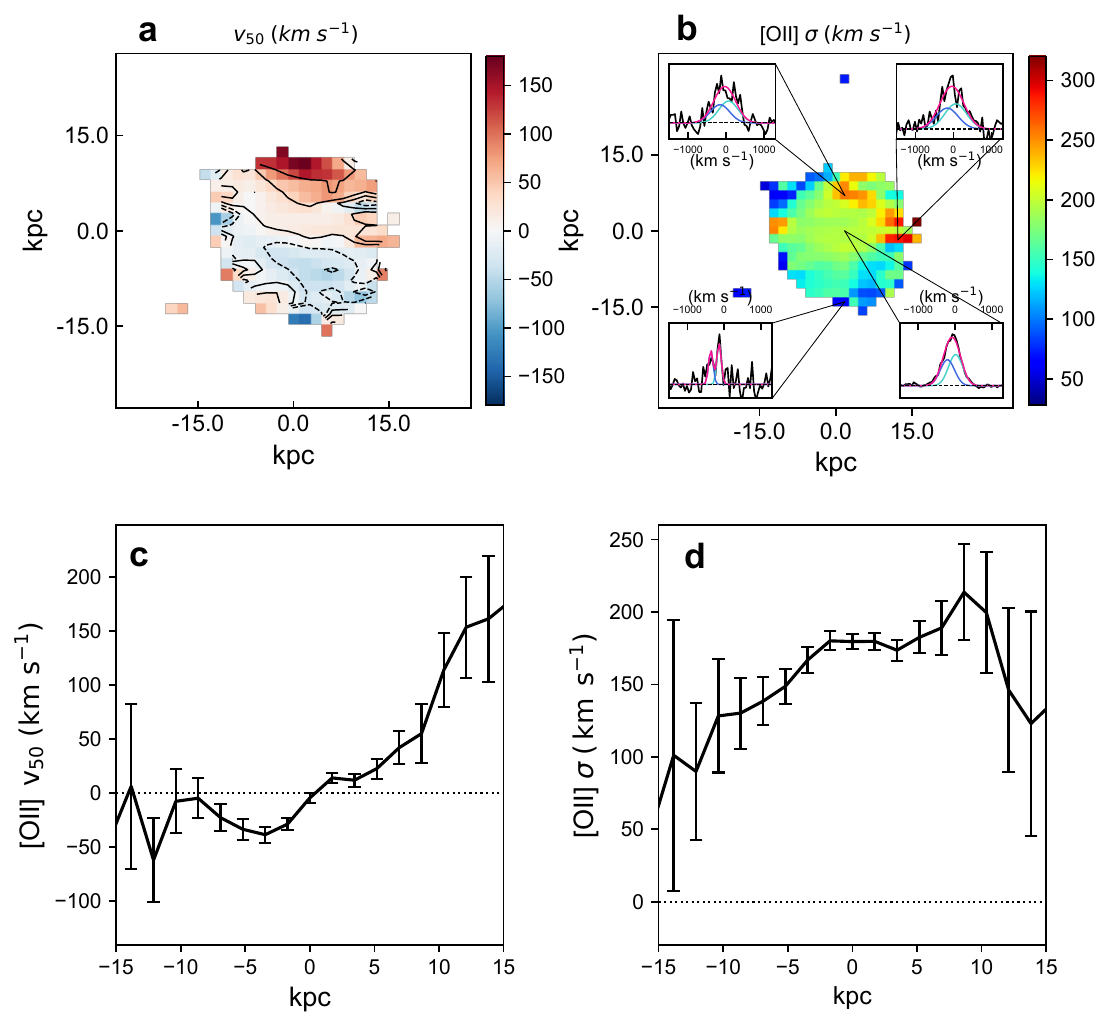}
  \end{center}
  \caption{\small {\bf Kinematics of the \oii \ emission. }
    (a) Velocity map of the \oii \ emission showing the central velocity ($V_{50}$) of
    the ionized gas relative to the systemic redshift of the source, $z=0.4512$.
    Contours mark where the central velocity is -50, -25, 0, 25, 75 \kms.
    A clear
    velocity gradient is seen from north to south across the nebula. 
    (b) The velocity dispersion $\sigma$ of the \oii \ emission, showing
    a higher dispersion in the center, north, and west regions of the nebula.
    Insets show spectra at various locations around the nebula,
    highlighting areas with high and low velocity dispersion.  Blue and cyan
    lines in the insets show Gaussian fits to the individual \oii \
    emission lines.
    (c) A position-velocity diagram through the center of the \oii \
    nebula along the north-south direction, averaging across the central 5
    columns, with 1$\sigma$ errors.
    (d) The velocity dispersion through the center of the
    nebula along the north-south direction, averaging across the central 5
    columns, with 1$\sigma$ errors.
  }
  \label{oii_kinematics}
\end{figure}

\clearpage

\begin{figure}
  \begin{center}
  \includegraphics[width=0.95\textwidth]{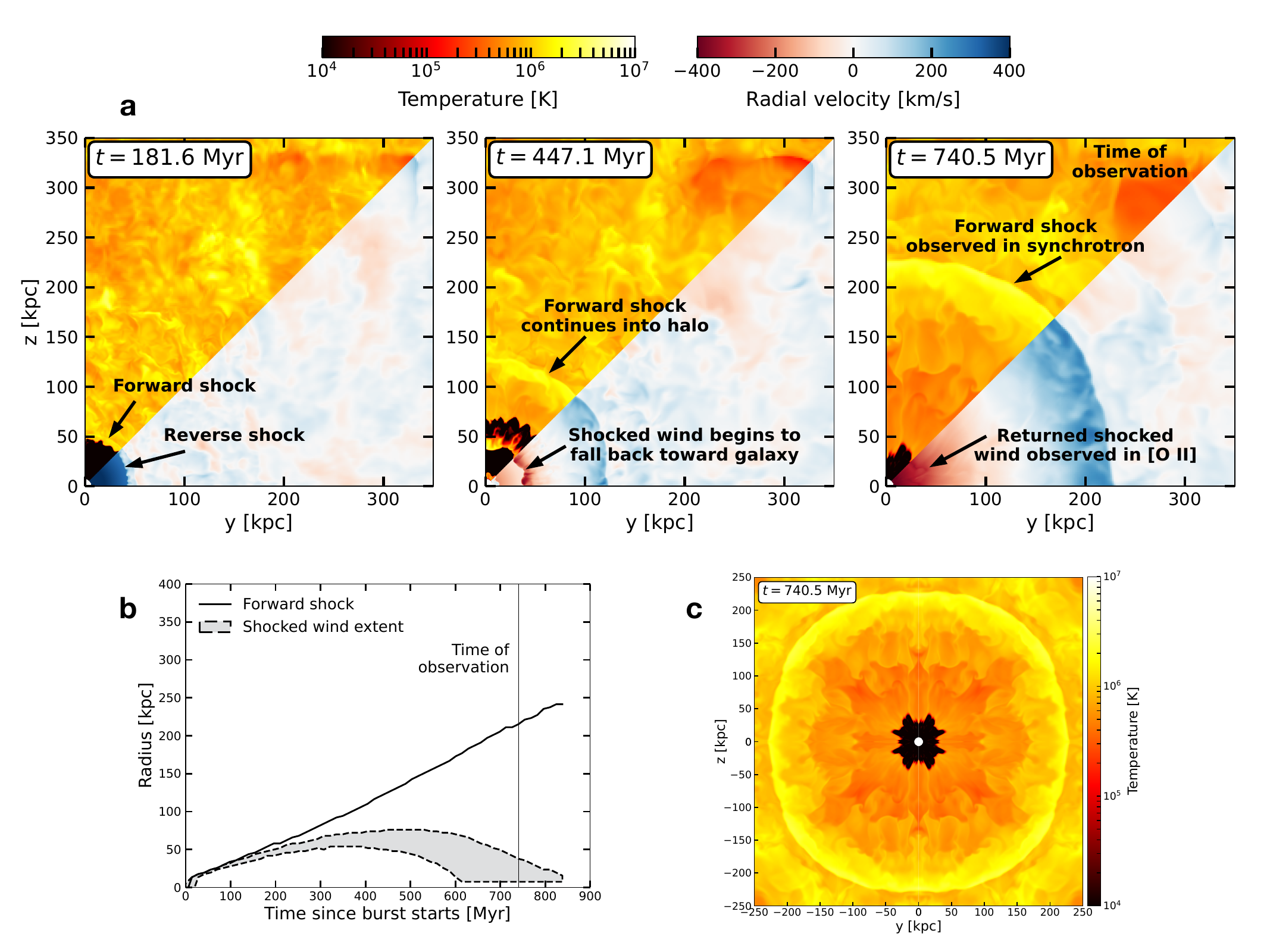}
  \end{center}
  \caption{\small {\bf Simulation of starburst-driven wind. }
    (a) Two-dimensional maps of gas temperature (upper left half of each panel) and gas
    radial velocity relative to the galaxy (lower right half of each panel), shown such
    that red colors indicate material falling onto the galaxy and blue colors indicate
    material outflowing from the galaxy, at three times since the onset of the starburst.
    The wind blows for 200 Myr, after which wind material that has been heated by the
    reverse shock falls back towards the galaxy, while the forward shock continues to
    expand into the halo. (b) Radius of the large-scale forward shock and the
    smaller-scale inner and outer boundaries of the shocked wind as a function of time
    since the onset of the starburst. (c) Spherical two-dimensional map (mirrored
    from the one quadrant simulated) of the gas temperature at the time of observation, 740
    Myr after the onset of the starburst and 540 Myr since the star formation and wind
    ceased.  The large yellow ring corresponds to where radio synchrotron
    emission is expected to arise, while the smaller-scale black region surrounding the
    central galaxy contains cool gas that could be observed as \oii \ emission.
  }  
\end{figure}

\setcounter{figure}{0}
\renewenvironment{figure}{\let\caption\edfigcaption}{}

\clearpage

\begin{methods}

\subsection{KCWI observations and data reduction.}

We observed ORC4 (RA=15:55:24.63 Dec=$+$27:26:34.3) with KCWI on the
Keck II telescope on March 27, 2022. The weather was clear and the
seeing was $\sim$0.9$\arcsec$; we observed at a position angle of 0
degrees and an airmass of 1.2.  We used the blue low dispersion (BL)
grating and the medium slicer on KCWI, which provides a spectral
resolution of R$=$1800, a spaxel size of 0.29$\arcsec$ $\times$
0.69$\arcsec$, and a FOV of 16$\arcsec$ $\times$ 20$\arcsec$ per
pointing. We configured the instrument to a central wavelength of 4700
\AA\ and used a detector binning of 2 $\times$ 2, which provides
wavelength coverage of 3700 $-$ 5700 \AA.  We observed a total of
three KCWI pointings, each spanning 16$\arcsec$ $\times$ 20$\arcsec$,
integrating for 20 minutes per pointing. One pointing was
centered on ORC4; these are the data we report here.  Two additional
pointings were centered 18$\arcsec$ to the north and to the south of
the first pointing; no emission was detected at either offset
pointing.  In total we observed a field of view of 16\arcsec
\ $\times$ 56\arcsec.

We reduced the data using the IDL version of the KCWI Data Extraction
and Reduction Pipeline (v1.2.1) and the IFSRED IDL library
\cite{Rupke14}.  Sky subtraction was performed using the IDL pipeline
with manual selection of a sky mask region within each pointing.  The
standard star BD+33d2642 was used for flux calibration.  Following the
pipeline stages, we resampled the data onto 0.29$\arcsec$ $\times$
0.29$\arcsec$ spaxel grids using the routine IFSR\_KCWIRESAMPLE. The
resulting data cube has dimensions of 56 $\times$ 66 spaxels, covering
16.2$\arcsec$ $\times$ 19.1$\arcsec$.  Assuming a
$\Lambda$CDM Planck 2020 cosmology\cite{Planck20} the physical
dimensions of a single KCWI pointing correspond to 97 kpc $\times$ 114
kpc at $z$=0.4512.

\subsection{Spatially-integrated spectrum and line fluxes.}

In addition to having spectra at each spaxel in the full datacube, we
created three spatially-integrated spectra by
summing the observed flux in rectangular apertures spanning the
central 3$\times$3, 7$\times$7, and 15$\times$15 spaxels,
corresponding to physical scales of 5, 12, and 26 kpc,
respectively. We detect strong \oii \ 3726\AA \ and 3729\AA \ and weak
\mgii \ 2796\AA \ and 2803\AA \ and \neiii \ 3869\AA \ line emission
in the spatially-integrated spectra (Extended Data Fig.~1), which also
include continuum emission from stars in the central galaxy.  The \oii
\ doublet is blended in most spaxels, due to high velocity dispersion.
The \oii \ and \neiii
\ emission trace ionized gas; while \mgii \ is ionized gas, it is a
resonant transition to the ground state
and has an ionization state that overlaps with neutral hydrogen, such
that it traces neutral gas\cite{Chisholm20}.
 Both \oii \ and \mgii \ trace gas temperatures of $\sim10^4$ K.
The \oii \ doublet is observed at air wavelengths of 5407.2 and
5411.2\AA; the spectroscopic redshift of ORC4 is therefore $z=0.4512$.
The [\textrm{Ne}~\textsc{v}] 3426\AA \ emission line is not
detected.

The spatially-integrated spectrum spanning the largest scales is used
to determine integrated fluxes.  The flux of the \oii \ doublet,
corrected for Galactic extinction, is $8.8 \ (\pm0.1) \times 10^{-16}
\ erg \ s^{-1} \ cm^{-2}$.  The total \mgii \ flux is $0.64
\ (\pm0.13) \times 10^{-16} \ erg \ s^{-1} \ cm^{-2}$; the ratio of
the \oii \ to \mgii \ flux is 14.
The restframe EW of the \mgii \ emission is 7 \AA, and the
observed ratio of the blue to red \mgii \ emission doublet is 1.9,
close to the intrinsic flux ratio of 2.0, which implies that the gas
is optically thin\cite{Chisholm20} and not substantially impacted by
absorption.  The flux of the \neiii \ 3869 \AA \ line is 0.72
($\pm0.11) \times 10^{-16} \ erg \ s^{-1} \ cm^{-2}$.
The \mgii \ and \neiii \ emission are too weak to detect in spatially
resolved maps.
In the spatially-integrated spectrum the kinematics of the \mgii \ and
\neiii \ emission are consistent with the \oii \ emission, both in
terms of velocity centroid and width.

\subsection{Spectral cube analysis.}

We fit Gaussian profiles to the observed emission lines, fitting
\mgii, \oii, and \neiii \ in the spatially-integrated spectra and \oii
\ in each spaxel of the full datacube.  As the \oii \ doublet is blended,
we fix the \oii
\ flux ratio to be 1.2, corresponding to an electron density of
$\sim200 \ cm^{-3}$.  We did not tie the kinematics of the \mgii,
\oii, and \neiii \ lines together. The continuum level around each
line is determined locally using the flux within 40 \AA \ on either side of
the emission singlet or doublet.

We define $V_{50}$ as the central velocity of the emission, as
determined by the centroid of the Gaussian fit.  The velocity
dispersion $\sigma$ is also determined by the Gaussian fit, from which
the instrumental resolution is subtracted in quadrature. The ratio of
$V_{50}$ to $\sigma$ in ORC4 is low, with a median value of 0.19; this
is an order of magnitude lower than what is seen in star-forming
galaxies\cite{Green14}. The statistical errors on $\sigma$ range from
$\sim$10 \kms \ at the center of the nebula to $\sim$20 \kms \ at a
radius of 5 kpc and $\sim$65 \kms \ at a radius of 10 kpc.  The
systematic errors on $\sigma$ are subdominant; changing the \oii
\ flux ratio from extreme values of 1.0 to 1.5 results in median
differences to $\sigma$ of only $\sim$2 \kms \ and does not change our results.

The systemic redshift of $z=0.4512$ is defined by the centroid of the
observed wavelengths of the \oii \ emission in the largest
spatially-integrated spectrum.  The \oii \ EW is calculated by
comparing the total flux in the \oii \ doublet to the fit of the
continuum level around \oii, which is divided by $(1+z)$ to obtain a
restframe EW.

In Figs. 1 and 3 we show those spaxels for which the \oii \ flux has a
signal to noise ratio of 3 or greater. We calculate the surface
brightness radial profile using RADP in IDL and fit for the slope of
the \oii \ profile from $6-15$ kpc, beyond the region of the seeing
disk.  The errors shown in Fig. 2 are standard errors, calculated from
the variation in the surface brightness within that radial bin, and
the errors in Fig. 3 are the standard errors calculated from the
variation in the central velocity fit across the central five spaxels
of the nebula.

\subsection{Surface brightness profile and Sersic fit.}

The central $\sim$5 kpc of the radial surface brightness profiles are
not resolved due to the observational seeing; from 6-15 kpc the best fit power law
slope to the \oii \ profile is $-2.8$.
We fit the radial surface brightness profile of both the \oii \ nebula
and the stellar continuum with a Sersic profile, convolving with the
PSF of the observations.  The Sersic profile is:
\begin{equation}
I(r)= I_e \ exp\{-(1.9992n - 0.3271) \ [(r/r_e)^{(1/n)} - 1)]\}
\end{equation}
where $I$ is the surface brightness as a function of radius, $I_e$ is
the surface brightess at the half light radius, $r_e$, and $n$ is the
Sersic index, which reflects the amount of curvature in the profile.
We use CONVOL in IDL to convolve this profile with a Gaussian with a
FWHM of 0.9\arcsec to account for the seeing.  The PSF-corrected
best-fit Sersic index to the \oii \ emission yields $n=1$, while for the
stellar continuum the best fit is $n=2$.  We verified these results
using GALFIT with a Gaussian PSF.  The stellar emission in the ORC4
central galaxy is more extended than the \oii \ emission; the galaxy
half light radius is 4.8 kpc.

\subsection{\bf Stellar mass and age estimation.}

There is another galaxy near ORC4, as seen in Fig.~1.
Long-slit spectroscopy reveals that it is at the same redshift as ORC4
(D.~S.~N. Rupke et al., in preparation) and is 36 kpc away; these
galaxies are near enough to likely be interacting\cite{Norris21c}.
We estimate stellar masses and ages for ORC4 and its companion galaxy 
by fitting the spectral energy distributions (SEDs) using
\textsc{Prospector} \cite{Johnson21}, using a redshift of $z=0.4512$
for both galaxies (Extended Data Fig.~2).  \textsc{Prospector}
utilizes gridless Bayesian parameter estimation and incorporates
Markov chain Monte Carlo (MCMC) posterior sampling with \textsc{emcee}
to infer the properties of the underlying stellar populations using
Flexible Stellar Population Synthesis models
(FSPS)\cite{Conroy09}. For ORC4, we fit broadband SDSS
DR8\cite{Aihara11} \textit{ugriz} and \textit{WISE}\cite{Wright10} W1
and W2 photometry, and for the companion galaxy we fit SDSS
\textit{ugriz} photometry. We assume a Kroupa initial mass
function\cite{Kroupa01} and a power law attenuation curve with index =
-0.7.

For ORC4 we assume a parametric $\tau$-model star formation history
with a burst where the star formation rate is given as $SFR \sim \exp
(t_{age}/\tau)$, and we allow an AGN contribution.  The free parameters
in the model are the galaxy stellar mass and age ($t_{age}$), the
fraction of stellar mass formed
in the burst ($f_{burst}$), the fraction of the stellar age at which a starburst
occurred ($f_{age,burst}$), $\tau$, and
the fraction of the light that is due to an AGN.
All timescales have units of Gyr.  We set a log-normal prior
on the stellar mass with a mode of $\ln (M_{*}/M_{\odot}) = 23$ and
$\sigma = 15$, as well as a log-uniform prior on $t_{age}$ such that
$0.1 < t_{age}/\text{Gyr} < t_{universe} (z)$ where $t_{universe} (z)$
is the age of the universe at the given redshift of $z = 0.4512$.  For
the companion galaxy we similarly assume a parametric $\tau$-model
star formation history, but we do not include a burst or an AGN contribution.
Stellar mass and $t_{age}$ are the only free parameters in the model for the
companion galaxy, and we assume the same priors that we implemented
for ORC4. For both ORC4 and its companion galaxy the stellar
metallicity ($\log Z_{*}/Z_{\odot}$) is updated for each random draw
of $M_{*}$ within $\pm 0.1$ dex of the stellar mass-metallicity
relation\cite{Ma16}.

The SED fit to the companion galaxy reveals that it is massive, with
log $(M_*/M_{\odot}) = 10.60 \ (+0.10/-0.16)$, and old, with $t_{age}
= 7.0 \ (+1.2/-1.4)$ Gyr, similar to what is found for the ORC4 central galaxy.

\subsection{Radio luminosity conversion.}

To convert the observed radio flux of the central source in ORC4 to a
luminosity at a restframe of 200 MHz, we use the following equation:
\begin{equation}
L_{\nu} = 4 \pi D_L^2 (1+z)^{-(1+\alpha)} (\nu/\nu_{obs})^{(1+\alpha)} (\nu_{obs}/\nu) F_{\nu_{obs}}
\end{equation}
where $D_L$ is the luminosity distance, $\alpha$ is the observed
spectral index of ORC4 of $-0.92$, $\nu$ is the observed-frame
frequency of the radio luminosity we are deriving, here equal to
200 MHz $\times (1+z)$, $\nu_{obs}$ is the observed frequency of 325 MHz,
and $F_{\nu_obs}$ is the observed flux of 1.43mJy.  We use this same
equation to convert to a restframe 1.4 GHz luminosity.

\subsection{Ionized gas mass estimate.}

We use the \oii \ luminosity to estimate the mass of the warm ionized
gas in ORC4.  Following Pandya et al. (2017)\cite{Pandya17}, we assume
\oii/H$\beta=6$ to compute an H$\beta$ luminosity and then use the
following equation:
\begin{equation}
M_{H\beta} = 28.2 \times 10^8 L_{H\beta,43} \ n_{e,100}^{-1} \ \msun
\end{equation}
where $L_{H\beta,43}$ is the H$\beta$ luminosity in units of $10^{43}
\ erg \ s^{-1}$ and $n_{e,100}$ is the electron density in units of
100 cm$^{-3}$.  We assume $n_{e,100}$=1 to facilitate a comparison
with the results of Pandya et al. (2017). The ratio of \oii/H$\beta$
assumed is similar to the value found in long-slit spectroscopy of
ORC4 ($\sim$5.5, D.~S.~N. Rupke et al., in preparation).

\subsection{Alternative scenarios for ionized gas origin.}

There are several alternative scenarios for the origin of the ionized
gas in ORC4.  
The first is that it is associated with
the interstellar medium (ISM) of the central galaxy. 
The velocity
gradient in the ionized gas could be due to rotation of the galaxy;
many early-type galaxies have spatially extended, rotating \oii
\ emission\cite{Pandya17}.  However, the velocity gradient is
asymmetric and the radial extent and the velocity dispersion of the
\oii \ emission in ORC4 are unusually high.  From the \oii
\ luminosity we estimate a warm ionized gas mass of $3\times10^7$
\msun, which is over an order of magnitude higher than that found for
massive early-type galaxies\cite{Pandya17}.  Additionally, the
observed \oii \ EW is far higher than what is seen in similar
early-type galaxies, such that it is very unlikely that this ionized
gas in ORC4 is simply cool gas in the ISM of the central galaxy.

Another possibility is that ORC4 may have a bipolar outflowing wind
driven by the central AGN, creating the extended \oii \ emission.
``Red geysers'' are early-type galaxies that have bisymmetric EW maps
with strong emission lines that appear to trace large-scale
outflows\cite{Cheung16,Roy21}.  Red geysers account for 5-10\% of
local, early-type galaxies\cite{Roy18}.
AGN-driven outflows can result in ionized gas velocity gradients, and
the central galaxy in ORC4 does have an AGN, such that it is plausible
that it has an AGN-driven outflow.  However, such outflows are
typically not circular and radio AGN outflows in particular are jet
driven and collimated\cite{Blandford19}.  Additionally, the \oii \ EW
map of ORC4 does not display the collimated, bisymmetric observational
signatures of red geyser EW maps.  Finally, the \oii \ luminosity of
ORC4 is two magnitudes higher than what is expected for a typical AGN
with the same radio continuum luminosity (Extended Data Fig. 3), and the SED fit
to the optical and infrared emission finds that only 3\% of the light is from
an AGN component.  Thus another mechanism is required to explain the strong,
extended \oii \ emission, and the high luminosity and EW in
particular, in ORC4.

It is also unlikely that the \oii \ emission could result from a cooling
flow as ionized gas in cooling flows typically has a complex, and
often filamentary, morphology\cite{McDonald10}.  Additionally, in
order for a cooling flow to be sustained the circumgalactic medium
should be well settled and not have any disrupting events; the
presence of the large-scale radio emission argues against this.  This
galaxy also does not appear to be in a well-defined overdensity or
galaxy cluster \cite{Norris21c} where cooling flows might be expected.

\subsection{Outflowing galactic wind simulation.}

We run the spherically-symmetric wind simulation using the Athena hydrodynamics code.  The
galaxy is modeled as an isolated, spherical wind injection region of
radius $r=$ 7 kpc.  The gravitational potential is provided by a
static NFW halo of virial mass 10$^{13}$ \msun, which was chosen from
the low-redshift stellar mass to halo mass abundance matching relation
\cite{Girelli20} for a galaxy of stellar mass $\sim10^{11.2}$ \msun,
similar to the value estimated for ORC4's central galaxy. The
simulated circumgalactic medium is initialized to be in isothermal
hydrostatic equilibrium with $T = 10^6$ K, close to the virial
temperature for a halo of mass $10^{13}$ \msun, and with a gas density
near the wind injection region of $10^{-26} \ g \ cm^{-3}$. The
circumgalactic medium is initialized with random velocity and density
perturbations with amplitudes on the order of 0.2 times the virial
velocity and 0.3 times the density, respectively. The wind turns on
immediately at the beginning of the simulation and blows
spherically-symmetrically with a velocity of 450 km s$^{-1}$ and with
a mass outflow rate of 200 \msun yr$^{-1}$, and then the wind turns
off at t = 200 Myr. We run just one octant of the sphere to reduce
computational costs.  The simulation domain is a cube with side length
350 kpc where the wind injection region is located in one corner, at
the origin. There are three levels of static mesh refinement: within a
44 kpc cube cornered at the origin, and thus closest to the wind
injection region, the spatial resolution is 438 pc. The resolution
decreases to 877 pc out to the edges of a 175 kpc cube, and then again
to 1.75 kpc out to the edge of the simulation domain at 350 kpc.

In the simulation that most closely matches the observational inputs, 
at 740 Myr after the starburst
begins, or 540 Myr since the star formation and resulting wind have ceased
(as the wind blows for 200 Myr),
the forward shock is located at 215 kpc in radius and has a
Mach number of 1.3, consistent with the shock Mach number required to
produce the observed synchrotron emission in ORCs\cite{Dolag23}.
At this same time, the previously-shocked wind
that gives rise to the \oii \ emission is falling back toward the
galaxy and extends from a radius of 0 to 35 kpc. The velocity
dispersion of the previously-shocked wind gas that is returning to the
galaxy is $\sim$120 km s$^{-1}$ within a 20 kpc radius, somewhat lower
than the \oii \ velocity dispersion observed in ORC4.  The velocity
dispersion in this simulation is a lower limit on the expected
velocity dispersion of \oii \ emitting gas, as the simulation is not
resolving the chaotic series of shocks that are likely producing the
\oii \ emission within the simulated $10^4$ K shocked wind region.
The
parameters of the wind (velocity, mass outflow rate, the timescale
that the wind is ``on") and the halo (virial mass, temperature,
density) were chosen to produce results that match the properties of
ORC4, though we found that a wide range of parameters were generically
able to produce a forward shock on large scales and shocked wind
falling back to the galaxy on small scales that could be observed
hundreds of Myr after the wind from the galaxy had ceased.  The
simulation does not predict observed \oii \ emission line luminosities
or the ionized gas mass in the shocked wind gas due to the computational
difficulty of radiative transfer modeling.

\subsection{Data and Code Availability.}

Raw data generated at the Keck Observatory are available at the Keck
Observatory Archive (koa.ipac.caltech.edu) following the standard 18-month
proprietary period after the date of observation.
The reduced KCWI spectral datacube and the results of the \oii \ emission
line fits are available on Zenodo at https://doi.org/10.5281/zenodo.8377942.
The code used is available upon request from the first author.

\end{methods}

\begin{addendum}
\item[Acknowledgments] We thank Fabrizio Brighenti for useful feedback on an
  earlier draft of the paper, and we thank Ray Norris and Huib Intema for
  sharing the GMRT discovery image of ORC4. 
  A.L.C. acknowledges support from the Ingrid and Joseph W. Hibben chair
  at UC San Diego. C.L. thanks Charles Cimino IV for computing resources and assistance.
  The data presented herein were obtained
  at the W. M. Keck Observatory, which is operated as
  a scientific partnership among the California Institute of
  Technology, the University of California, and the National
 Aeronautics and Space Administration. The Observatory
  was made possible by the financial support
  of the W. M. Keck Foundation. The authors wish to recognize
  and acknowledge the very significant cultural role
  and reverence that the summit of Maunakea has always
  had within the indigenous Hawaiian community. We are
  most fortunate to have the opportunity to conduct observations
  from this mountain.
\item[Author Contributions]
  D.S.N.R. and A.L.C. conceived the observations, following C.A.T.'s suggestion
  that ORCs may be similar to the galaxies we have been studying. A.L.C.
  obtained the observing time, S.P.
  performed the KCWI observations and led the data reduction.  A.L.C. performed
  all of the data analysis, with input from D.S.N.R.  C.L. ran the starburst-wind model
    simulation, and K.W. performed the SED fitting. A.L.C. wrote the manuscript, with input from all coauthors.  Figures were created by S.P., A.L.C., K.W., C.L., and R.H.
\end{addendum}

\clearpage

\begin{figure}
  \begin{center}
  \includegraphics[width=1.0\textwidth]{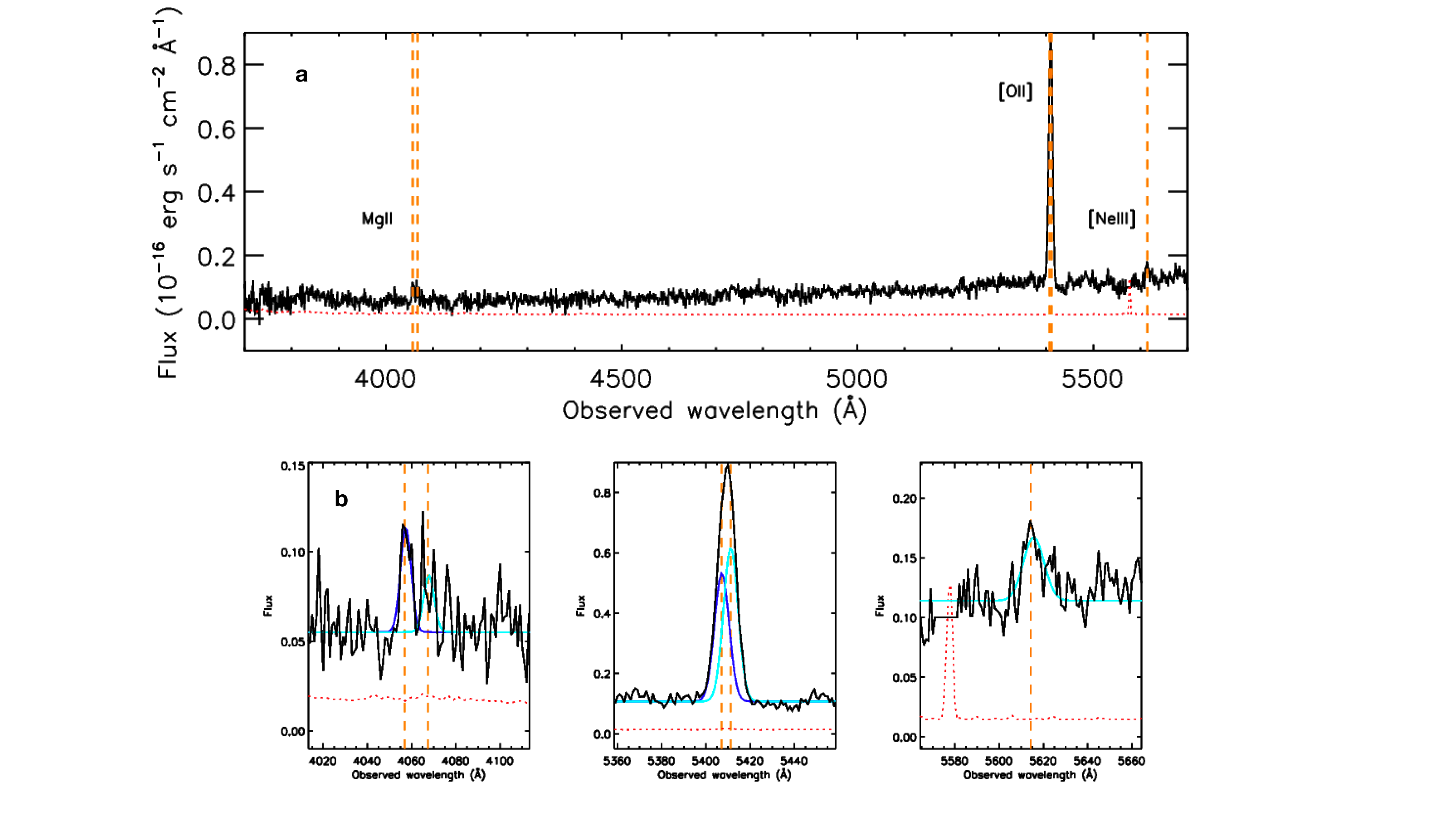}
  \end{center}
  \caption{\small {\bf Spatially-integrated spectrum of the gas nebula and host galaxy
      stellar continuum emission in ORC4.} (a) Strong \oii \ 3726, 3729 \AA \ 
    and  weak \mgii \ 2796, 2803 \AA \
    and \neiii \ 3869 \AA \ emission
    are observed in a spatially-integrated spectrum of ORC4
    (spanning the inner 26x26 kpc), at
    wavelengths that correspond to $z=0.4512$; weak stellar continuum is also
    detected. The solid black line shows the observed spectrum, the pink dotted
    line is the 1$\sigma$ error spectrum,
    and the orange dashed vertical lines indicate the observed wavelengths of
    \mgii, \oii, and \neiii \ at the spectroscopic redshift of the source.
    (b) Gaussian fits to
    the \mgii \ and \oii \ emission doublets and the \neiii \ singlet in the
    spatially-integrated spectrum.
    Blue and cyan lines show fits to the individual emission lines in each
    doublet.}
  \label{integrated_spec}
\end{figure}

\begin{figure}
  \begin{center}
  \includegraphics[width=\textwidth]{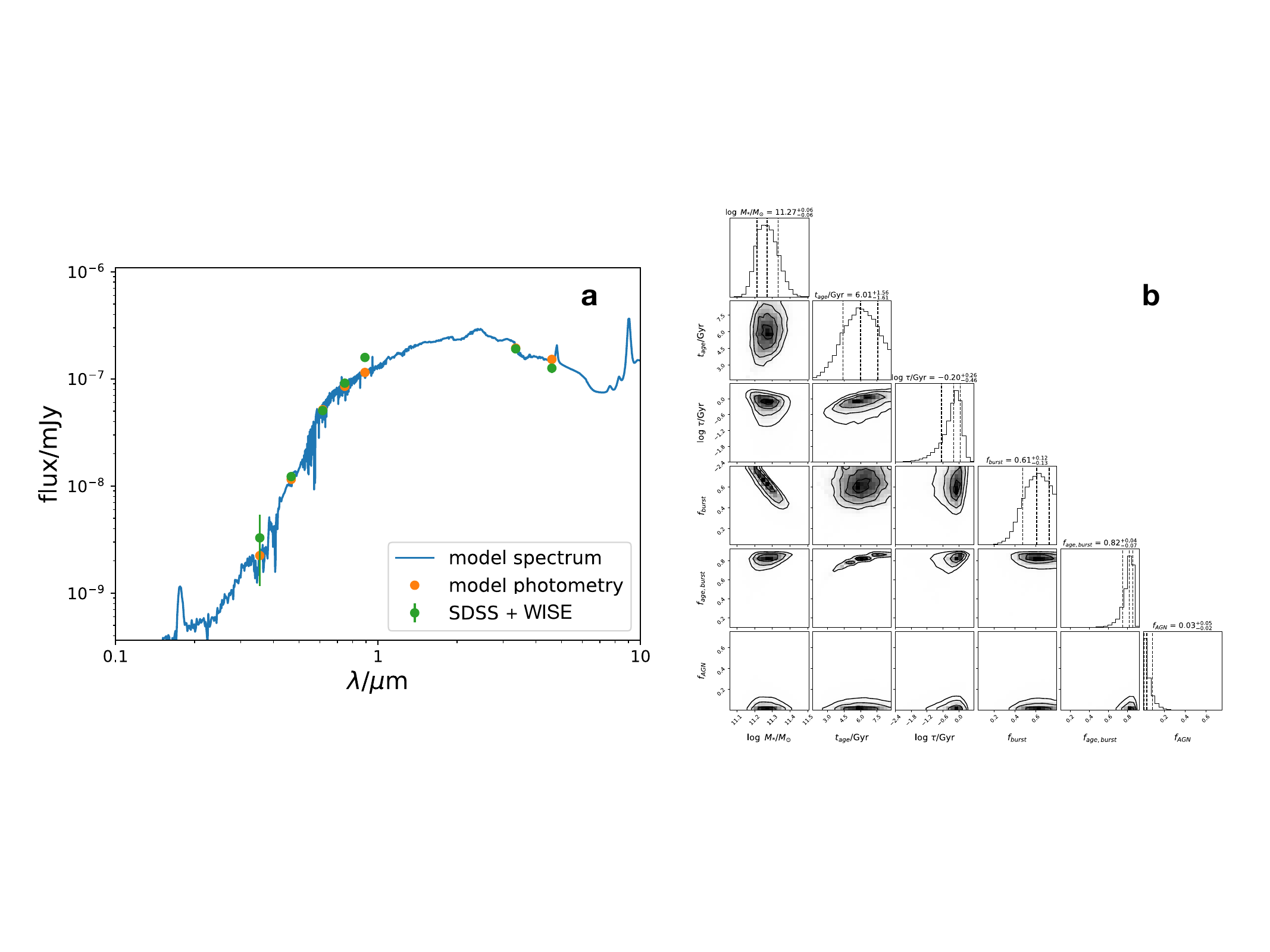}
  \end{center}
  \caption{{\bf Spectral energy distribution (SED) fit to optical SDSS and near infrared WISE 
    photometry of ORC4.} (a) The observed photometry of ORC4 is shown as green circles with 1 $\sigma$ error bars, while orange circles show the photometry implied from the best fit stellar population model, including an AGN contribution, which is shown in blue. Flux values are given in mJy and observed frame wavelengths in $\mu$m. (b) The derived distributions from the SED fit for the stellar mass, stellar age, star formation history parameters, and AGN contribution for ORC4, as well as the covariance between the parameters.}
\end{figure}

\begin{figure}
  \begin{center}
  \includegraphics[width=0.85\textwidth]{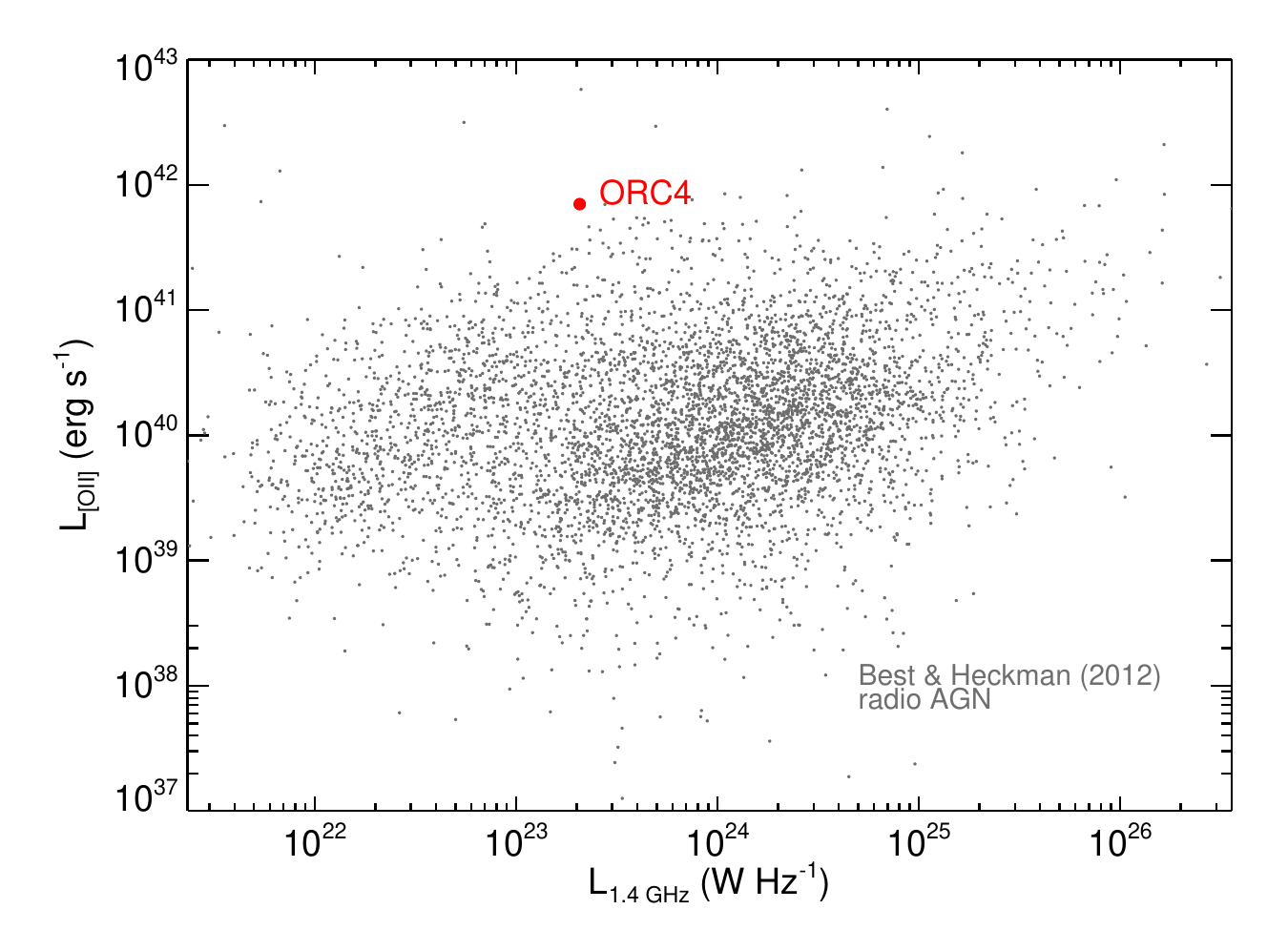}
  \end{center}
  \caption{{\bf Comparison of ORC4 radio continuum and \oii \ line luminosity to radio AGN.} The 1.4 GHz radio continuum luminosity and \oii \ emission line luminosity of ORC4, shown with a red circle, compared to the radio-loud AGN sample of Best \& Heckman (2012)\cite{Best12}, shown with grey points. The \oii \ luminosity of ORC4 is two orders of magnitude higher than the median value for an AGN with the same 1.4 GHz radio continuum luminosity. }
\end{figure}

\end{document}